\documentclass[12pt,twoside]{article}
\RequirePackage{xspace}
\usepackage[final]{graphics}
\usepackage{relsize}
\usepackage{epsfig,color}
\usepackage{graphicx}

\def\babar{\mbox{\slshape B\kern-0.1em{\smaller A}\kern-0.1em B\kern-0.1em{\smaller A\kern-0.2em R}}}
\def\Bbar    {\kern 0.18em\overline{\kern -0.18em B}{}\xspace}
\def\BB      {\ensuremath{B\Bbar}\xspace}
\newcommand{\mev}{\ensuremath{\mathrm{\,Me\kern -0.1em V}}\xspace}
\newcommand{\mevcc}{\ensuremath{{\mathrm{\,Me\kern -0.1em V\!/}c^2}}\xspace}
\def\B       {\ensuremath{B}\xspace}
\def\Dstarp  {\ensuremath{D^{*+}}\xspace}
\def\DeltaE     {\mbox{$\Delta E$}\xspace}
\def\en         {\ensuremath{e^-}\xspace}   
\def\ep         {\ensuremath{e^+}\xspace}
\def\Bub     {\ensuremath{B^-}\xspace}
\def\Dp      {\ensuremath{D^+}\xspace}
\def\Bm      {\ensuremath{\Bub}\xspace}
\def\Y#1S{\ensuremath{\Upsilon{(#1S)}}\xspace}
\def\FourS {\Y4S}
\def\to      {\ensuremath{\rightarrow}\xspace}
\begin{document}

\title{Study of $B^{-} \rightarrow D_{2}^{*}(2460)^{0} \pi^{-}$ and $B^{-} \rightarrow D_{1}(2420)^{0} \pi^{-} $ Decays}

\author{Vance O. Eschenburg}

\maketitle

\begin{center}
University of Mississippi at Oxford \\
University, MS 38677 USA\\

\vspace{+5mm}

{\em Representing the \babar{} Collaboration} \\

\vspace{+5mm}

Stanford Linear Accelerator Center \\
Stanford University \\
Stanford, CA 94309 USA \\

\vspace{+10mm} 

{\em Presented at the 2004 Meeting of the Division of Particles and Fields }\\
{\em of the American Physical Society }\\
{\em Riverside, CA USA }\\
{\em August 26, 2004  - August 31, 2004 }\\ 
{\em Submitted to International Journal of Modern Physics A }
\end{center}

\begin{abstract}
We report on a study of $B$ mesons decaying into one of the narrow P-wave charm resonances, $D_2^{*}(2460)^{0}$ and $D_1(2420)^0$. Our preliminary results are based on 89 million \BB pairs collected with the \babar~detector at the PEP-II asymmetric $B$ Factory. Our study will be useful in the investigation of the properties of Heavy Quark Effective~Theory. 
\end{abstract}

\section{Introduction}
\begin{figure}[t]
\begin{center}
\includegraphics[scale=0.385]{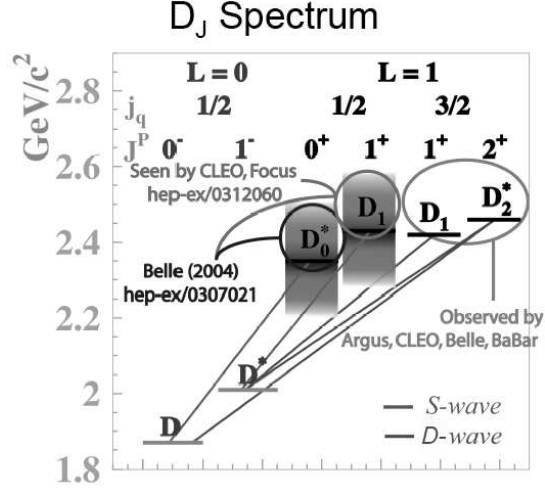}
\vspace{-3mm}
\caption{Spectroscopy of the $D_{J}$ mesons.}
\end{center}
\end{figure}

The notation $D_{J}$ ($D^{**}$) refers to orbitally excited $D$ mesons, consisting of a charm quark and an up or down anti-quark with an orbital angular momentum of $L=1$ \hbox{(P-wave)}. Measurement of the properties of these particles tests theories such as Heavy Quark Effective Theory (HQET)\cite{Isgur}. HQET predicts two sets of doublets\cite{Rujula} (Fig.~1), one with $j= \frac{3}{2}$ ($j$ being the sum of the spin of the lighter quark and the orbital angular momentum) and another with $j=\frac{1}{2}$. The four neutral $D_{J}$ mesons are $D_{0}^{*}(j=\frac{1}{2})^{0}$, $D_{1}(2420)^{0}$, $D_{1}(j=\frac{1}{2})^{0}$, and $D_{2}^{*}(2460)^{0}$ (Table 1). The two $D_{J}$ with $j=\frac{1}{2}$ are broad resonances (with a width of a few hundred $\mev$) while the other two states are narrow resonances (width $20-30 \mev$)\cite{Belle}$^{,}$\cite{PDG}.

\begin{table}
\null
\hskip+5mm
\footnotesize
\begin{tabular}{c|cccc}
\raisebox{3pt}[10pt][0pt]{State}  &  \raisebox{3pt}[10pt][0pt]{ $D^{*}_{0}(j=\frac{1}{2})^{0}$} & \raisebox{3pt}[10pt][0pt]{$D_{1}(2420)^{0}$} & \raisebox{3pt}[10pt][0pt]{$D_{1}(j=\frac{1}{2})^{0}$} & \raisebox{3pt}[10pt][0pt]{$D^{*}_{2}(2460)^{0}$}  \\ 
\hline
\rule{0pt}{10pt} $J^{P}$      & $0^{+}$ 	& $1^{+}$               & $1^{+}$                   &    $2^{+}$      \\ 
 Mass ($\mevcc$) & $ 2308\pm17\pm15\pm28$ & $2422.2\pm1.8$            & $2427 \pm 26 \pm 20 \pm 15 $    &  $2458.9 \pm 2.0$  \\
 Width ($\mev$)    	& $276\pm21\pm18\pm60$  & $18.9^{+4.6}_{-3.5}$         & $ 384^{+107}_{-75} \pm 24 \pm 70$  & $23 \pm 5$  \\
 Decays        		& $D \pi$                 &  $D^{*} \pi $  & $D^{*} \pi $                   & $D \pi$, $D^{*} \pi$ \\ 
 
\end{tabular}
\caption{Properties of  $D_{J}^{0}$ mesons. Mass and width measurements from Ref. 3 and 4.}
\end{table}

Of special interest is the ratio of the two branching fractions of the narrow states: $ R \equiv \frac{{\cal B}(B^{-}\rightarrow D_{2}^{*}(2460)^{0}\pi^{-})}{{\cal B}(B^{-}\rightarrow D_{1}(2420)^{0}\pi^{-})}$. The CLEO Collaboration measured this ratio\cite{CLEO} to be $1.8 \pm 0.8$ while the Belle Collaboration's value\cite{Belle} is $0.77 \pm 0.15$\null. One theoretical study\cite{Leibovich} uses HQET to calculate this value's range to be $0.0<R<1.5$.  Another prediction\cite{Neubert} narrows this value down to $R \approx 0.35$.

\section{Analysis Method}
The data, 89 million \BB pairs at the \FourS resonance,  were collected by the \babar{} detector\cite{b1} using the PEP-II asymmetric-energy $B$ Factory\cite{b2} at the Stanford Linear Accelerator Center. The $B^{-}$ mesons are reconstructed via the final state $D^{*+}\pi^{-}\pi^{-}$ or $D^{+}\pi^{-}\pi^{-}$ for studying the properties of the $D_{J}$ resonances. $D^{*+}$ candidates are reconstructed with a $D^{0}$ and a $\pi^{+}$. $D^{0}$ candidates are reconstructed as $D^{0} \rightarrow K^{-}\pi^{+}, K^{-}\pi^{+}\pi^{0},K^{-}\pi^{+}\pi^{-}\pi^{+}, K_{S}^{0}\pi^{+}\pi^{-}$, and the $D^{+}$ candidates as $D^{+}\rightarrow K^{-}\pi^{+}\pi^{+}$ , $K_{S}^{0}\pi^{+}$. Monte Carlo events are used for the optimization of selection criteria and estimation of the reconstruction efficiency. Tight particle identification requirements ensure that clean kaons and pions are used in the reconstruction of the $B$ mesons.

The inclusive branching fractions of $ \B^{-} \to D^{+} \pi^{-} \pi^{-} $ and $ \B^{-} \to \Dstarp \pi^{-} \pi^{-} $ are extracted from fits to $ \DeltaE = E^*_B - E^*_{\rm beam}$. $E^{*}_{\rm beam}$ is the center-of-mass~(CM) energy of the \ep/\en beam. $E^*_B$ is the CM energy of the \Bm~candidate. The signal is described by a Gaussian and the background by a linear function.

For the measurement of the exclusive branching fractions of the narrow resonances, the mass distributions $m(\Dstarp\pi)$ and $m(\Dp\pi)$ are fitted (Fig.~2).  The signal is described by Breit-Wigner functions convolved with Gaussians that take the detector resolution into account. The broad resonances and the background from non-resonant $\B \to D^{(*)} \pi \pi$ are represented by relativistic Breit-Wigner functions. The shape of the combinatorial background is obtained from events with \DeltaE outside the signal region.  Additional details of this analysis may be found in Ref.~10.

\begin{figure}
\null
\hspace{+5.0mm}
\includegraphics[scale=0.303]{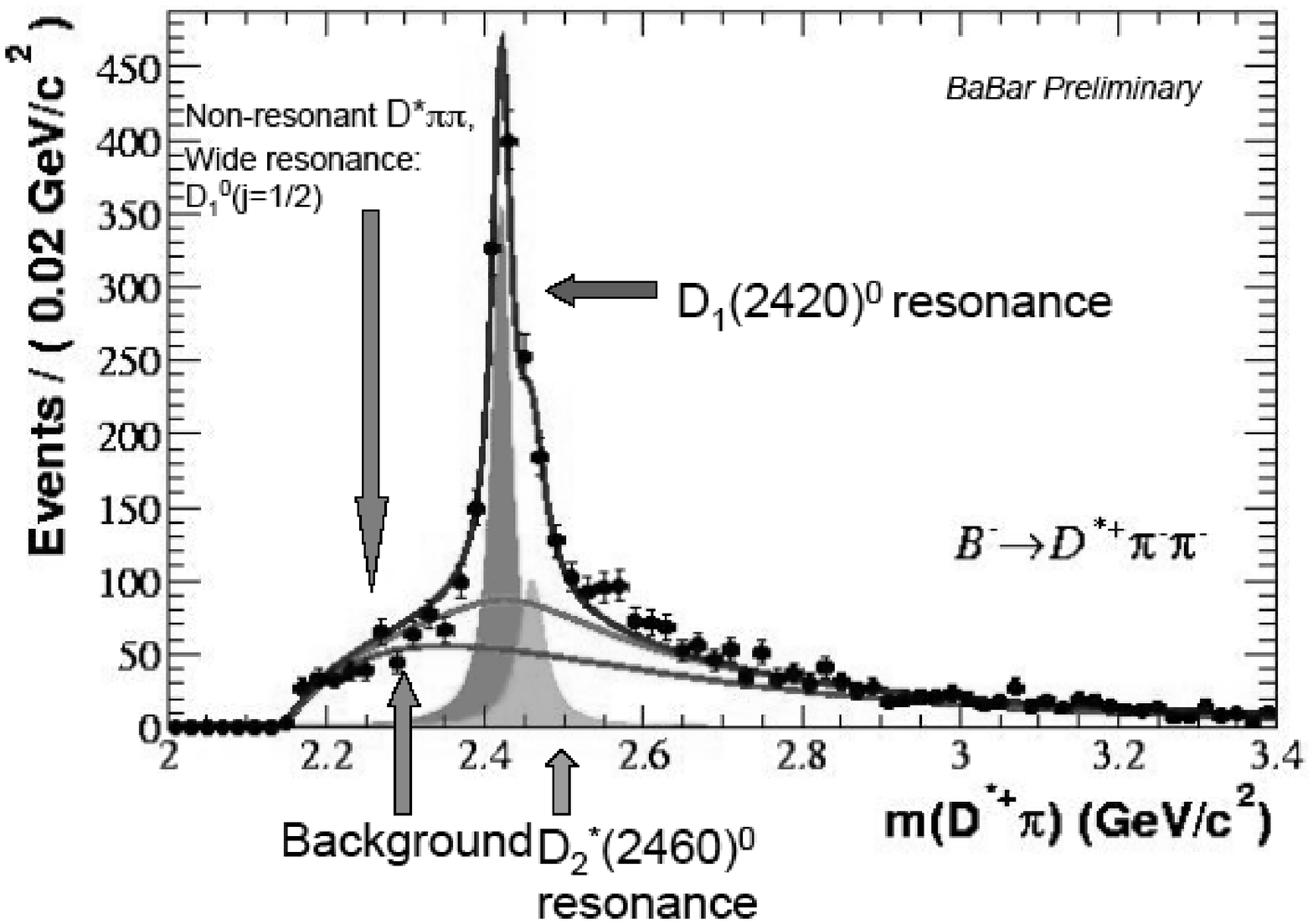}
\hspace{+5.0mm}
\includegraphics[scale=0.299]{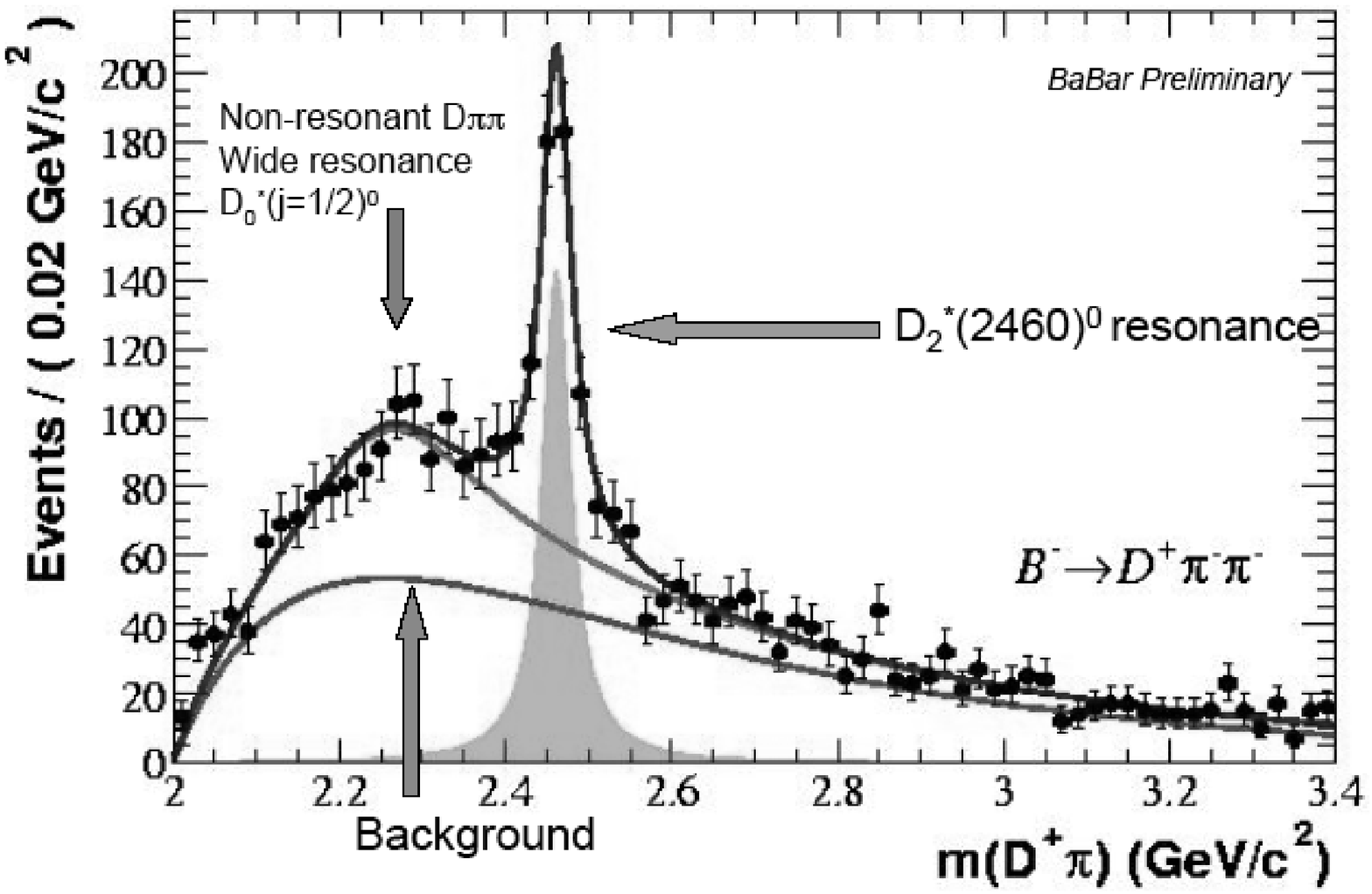} \\
\begin{center}
\vspace{-5mm}
\caption{{\it Left:} Distribution of $m(D^{*} \pi)$ with the narrow resonances $D_{1}(2420)^{0}$ and $D_{2}^{*}(2460)^{0}$. {\it Right:} Distribution of $m(D \pi)$ with narrow resonance $D_{2}^{*}(2460)^{0}$. {\it Both:} The lower background line represents combinatorial background, while the upper background line also contains the wide resonant states and non-resonant $B^{-} \to D^{(*)+} \pi^{-} \pi^{-}$ events.}
\end{center}
\vspace{-5mm}
\end{figure}

\begin{table}[b]
\hspace{+5mm}
\begin{tabular}{cc}
Mode & Branching Fraction ($\times 10^{-3}$) \\
\hline
 $ B^{-} \rightarrow D^{*+}\pi^{-}\pi^{-} $ & $ 1.22 \pm 0.05 \pm 0.18 $ \\
   $ B^{-}\rightarrow D^{+}\pi^{-}\pi^{-} $ &  $ 0.87 \pm 0.04 \pm 0.13$  \\
     $(B^{-} \rightarrow D_{2}^{*}(2460)^{0}\pi^{-})\times(D_{2}^{*}(2460)^{0} \rightarrow D^{+} \pi^{-} ) $ & $0.29 \pm 0.02 \pm 0.05$  \\
       $(B^{-}\rightarrow D_{1}(2420)^{0}\pi^{-})\times(D_{1}(2420)^{0}\rightarrow D^{*+} \pi^{-} ) $ & $0.59 \pm 0.03 \pm 0.11$ \\
         $(B^{-}\rightarrow D_{2}^{*}(2460)^{0}\pi^{-})\times(D_{2}^{*}(2460)^{0}\rightarrow D^{*+} \pi^{-} ) $ & $0.18 \pm 0.03 \pm 0.05$ \\
	 \end{tabular}
	 \caption{Preliminary branching fractions.}
	 \end{table}

\section{Results and Acknowledgment}

All results are preliminary. Table 2 lists the results for the inclusive and exclusive branching fractions. Our measurements of the branching fractions and of the branching ratio $R = 0.80\pm0.07\,\hbox{(stat.)}\pm0.16\,\hbox{(syst.)}$ agree with the results\cite{Belle} of the Belle Collaboration.  

The~author thanks the \babar~Collaboration, the SLAC accelerator group
and all contributing computing organizations. He~was supported
 by U.S. Dept.~of Energy grant DE-FG05-91ER40622.


\end{document}